# Mobile Apps for Children's Health and Wellbeing: Design Features and Future Opportunities

Jamie Lee*, BS[1], Zhaoyuan Su*, MS[1], Yunan Chen, PhD[1]
[1]University of California, Irvine, Irvine, CA, USA

**Abstract**
*Mobile health apps hold great potential for promoting children's health and wellbeing. However, there is limited understanding of how these technologies are currently designed to support children with their health concerns or wellness goals. To gain insight into the current landscape of mobile apps designed for children's health, we retrieved and reviewed 43 apps from IOS and Google Play store that are specifically marketed for children. Our qualitative analysis identified the dominant health focuses and goals of children's mobile health apps. We analyzed the primary users and their expectations as well as the methods of engagement and involvement adopted. Based on our findings, we discussed the opportunities to support children with chronic illnesses through mobile apps, design for dual use, and design for age appropriateness and digital health safety. This study provides insights and recommendations for app designers, health researchers, and policymakers on strategies for engaging children and parents while also promoting children's health and wellbeing through mobile technology.*

**Introduction**

With the increasing smartphones and tablets availability, mobile applications have become a popular tool for promoting health and wellbeing among children. These consumer health technologies have the potential to provide children with personalized and engaging content that can support their physical, emotional, and mental health in their daily lives. For example, apps have been developed to provide children with tools (e.g., illness diary) and resources (e.g., educational content) to improve their nutrition and physical activity levels or manage chronic conditions such as asthma or diabetes. They can also provide children with content to support their mental health, such as mindfulness exercises or coping strategies for managing stress and anxiety.

Unlike designing mobile health apps for adults, designing effective mobile apps for children's health and wellbeing presents unique challenges. One of the most important considerations is the developmental needs of children. Children's cognitive, social, and emotional development varies widely depending on age [1,2], and app designers need to take these differences into account when designing mobile health apps. For example, young children may not have their own smartphones, which may require more guidance and supervision from their caregivers or parents. Their cognitive development may also limit their understanding of comprehensive health data and other complex information. On the other hand, older children may be more capable of navigating complex interfaces, design features, and healthcare language. Another factor to consider is the potential privacy and safety risks associated with digital technology. Children are particularly vulnerable to interactive mobile health apps, due to their still-developing mental model of safety and privacy concerns [3,4]. In addition, there is a need to consider the potential impact of mobile apps on children's social and emotional development. For example, excessive screen time and social media use have been associated with negative health outcomes such as poor sleep quality and anxiety [5]. To address these challenges, an increasing number of researchers are studying how mobile health apps can support children's health.

In fact, the health informatics community has long been interested in studying and designing mobile health technologies to support children's health and wellbeing. For instance, researchers have examined the effectiveness of mobile apps in improving children's physical activity levels [6], nutrition [7], sleep quality [8], and mental wellness [9]. They have also explored how app design can be tailored to children's developmental needs and preferences, by adopting gamification [10,11], storytelling [6,12], and interactive features (e.g., customized health data stickers [13–15]) to engage and motivate children.

While previous research has provided valuable insights into designing mobile health applications to promote children's health and well-being, there is still limited understanding of how commercial health apps are designed to support children and their caregivers. Gaining a deeper understanding of this is critical to identify underserved health domains, existing strategies to engage children and parents, areas for improvement, and future design opportunities. This paper investigates the following questions: 1) what are the health focuses and goals of children's mobile health apps? 2) who are the users of the children's health apps and their expected roles? 3) how are users engaged or involved in the app through design? To answer these questions, we first retrieved and qualitatively analyzed 43 mobile apps that are specifically designed for children's health from iOS Apple, and Android Google Play stores. Then we categorized the apps into four health focuses and identified five main goals. Although these are apps marketed for children, we discovered that there is a duality of user roles: both children and parents/caregivers are the expected users. Our analysis sheds light on potential design advancements in mobile apps for chronic health conditions in children, improved collaboration between parents/caregivers and children, and the importance of age-appropriate design and digital health safety.

---

* These two authors contributed equally to this work

**Methods**

We conducted a qualitative study to gain insight into the landscape of commercial apps created for children's health and wellbeing. Qualitative methods are ideal for examining the design and use of health technology and revealing the attitudes, stories, and beliefs that drive human behavior[16,17]. Our study involved a review and analysis of the app features, the role of users, user engagement strategies, and a collection of user reviews.

*App Selection*

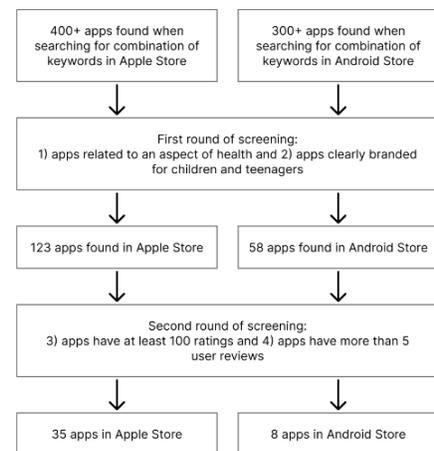

Fig 1. App Selection

To gather a list of currently used and relevant apps for further analysis, in October 2022, a systematic search was conducted on the iOS Apple App Store and the Android Google Play Store. We searched for apps using a combination of one child-related keyword ("kid", "children", "teen", or "teenager") and one health-related keyword ("health", "wellness", "asthma", "cystic fibrosis", "diabetes", "obesity", "ADHD", or "Autism"). The use of "health" and "wellness" as health-related keywords was intended to encompass the physical, mental, and social dimensions of well-being. Additionally, several prevalent chronic childhood illnesses were incorporated[18], as they constitute a crucial aspect of health.

The initial search queries resulted in over 400 apps. To identify apps that are relevant, we first used the following eligibility criteria: 1) the app is related to an aspect of health (physical, mental, or social), and 2) the app is clearly branded for children and teenagers. After removing duplicate apps from the initial screening, the searches returned a total of 181 apps consisting of 123 from the Apple store and 58 from the Android store. To exclude poorer quality apps, we used both ratings and reviews as criteria for measuring popularity. Since not all consumers who rate an app leave a review, using both ratings and reviews may result in more accurate indications of quality and usage. Two eligibility criteria were added: 3) the app contains at least 100 ratings, and 4) the app lists more than five user reviews. We screened the 181 apps using these criteria, and the second round of screening resulted in 35 apps from the Apple store and eight apps from the Android store. A total of 43 apps were retrieved and further analyzed in this study (see Fig. 1).

*Data Collection and Analysis*

To analyze the 43 children's health apps, the app store page content, app features, and user reviews were extracted to understand the main health focuses, the differences in children and parent roles in the apps, and the users' perception towards the apps. A codebook was created iteratively through discussion with the three authors over a period of eight weeks. The health focus, age group, and main objective of the app were assessed using direct quotes and explanations from the developer in the app descriptions. Additionally, for user roles, we deduced who the target users are, expected user actions, and engagement strategies. Two authors independently coded all the apps and compared their results using constant comparison methods[19]. Any disagreements were resolved through discussion. The three researchers met on a weekly basis to discuss the findings and analysis.

All 43 apps were downloaded, sample information was entered, and app features were recorded. When applicable, the same sample information was submitted to understand app output. Two authors analyzed these data iteratively through multiple rounds of comparison using the codebook and convened weekly to discuss and resolve any differences. In addition to reviewing the features of the mobile apps, we analyzed user reviews for the apps selected. For our study, user reviews provided feedback about users' perceptions of children's mobile health apps, their expectations, and their experiences. We randomly sampled and analyzed 200 reviews that were posted within the last five years to keep the content relevant since many apps have gone through updates and older reviews may no longer be applicable. During the coding process, the two authors iteratively evaluated the reviews and met with the third author for discussion. After, findings from all datasets were summarized and all authors met to resolve disagreements.

**Results**

In this section, we first summarize the 43 apps collected and identify the health focus of each app (see Table 1). Then we analyze children's and their parents' roles in managing healthcare and the expected work for them through the apps. Lastly, we report on the mechanisms designed in the apps to engage and involve users.

**Table 1: Summary of Apps and their Health Focus**

| Health Focuses | Subcategories | Apps |
|---|---|---|
| Self-care & management (25) | Overall wellbeing (5), mental wellbeing (10), physical fitness (4), oral hygiene (5), menstrual cycle (1) | *Apple:* 1) Habitz: Kids Learn Good Habits, 2) GoNoodle - Kids Video, 3) Cosmic Kids, 4) The Happy Child-Parenting App, 5) Smiling Mind, 6) Moshi Kids: Sleep & Meditations, 7) Breathe, Think, Do with Sesame, 8) Meditation for Kids, 9) Children's Sleep Meditations, 10) Mindful Powers, 11) Calm Kids: Mindfulness & Yoga, 12) Storybook - Bonding & Bedtime, 13) Baby Mozart - Children Music, 14) New Horizon: Sleep Meditations, 15) GoNoodle Games, 16) Kids morning exercises, 17) Toothy: Teeth Brushing Timer, 18) Brusheez - The Little Monster Toothbrush Timer, 19) Philips Sonicare for Kids, 20) Disney Magic Timer by Oral-B, 21) hum kids by Colgate, 22) Teen Period Tracker<br>*Android:* 1) Happy Kids Timer Chores, 2) Exercise for Kids at Home, 3) Yoga for Kids & Family fitness |
| Developmental health (6) | - | *Apple:* 1) Huckleberry: Baby & Child, 2) Growth, 3) CDC's Milestone Tracker, 4) BabySparks - Developmental App<br>*Android:* 1) Kinedu: Baby Development Plan, 2) Child Growth Tracker |
| Medical services (5) | General medical conditions (4), mental wellbeing (1) | *Apple:* 1) Children's Health VirtualVisit, 2) Texas Children's Anywhere Care, 3) Anytime Pediatrics, 4) MyCookChildren's, 5) Teen Counseling<br>*Android:* N/A |
| Children w/ special needs (7) | - | *Apple:* 1) Otsimo \| Special Education AAC, 2) Kids Autism Games - AutiSpark, 3) Otsimo \| Speech Therapy SLP, 4) Speech Blubs: Language Therapy<br>*Android:* 1) Language Therapy for Children, 2) Memory & Attention Training, 3) Card Talk |

*Health and Wellness Focus*

We identified four general aspects of children's health mobile apps: 1) self-care and management, 2) developmental health, 3) medical services, and 4) special needs. 25 out of 43 apps fall under the self-care and management category, and these apps can fit into several subcategories: overall wellbeing (five apps), mental wellbeing (10 apps), physical fitness (four apps), oral hygiene (five apps), and menstrual cycle (one app). The apps focusing on overall wellbeing offer services for physical, mental, and social health, such as meditations and exercise, while mental wellbeing apps are specifically for mental health, such as those that focus on anxiety. To give an example, GoNoodle - Kids Videos promotes physical and mental activities in the description by stating "*Make screen time active with 300+ dance videos, yoga exercises, and mindfulness activities for kids!*" while Smiling Mind advocates itself as a mental health app: "*Smiling Mind is a FREE mindfulness meditation app [...] mindfulness meditation is about mental health and looking after the mind.*"

The six developmental health apps concentrate on the milestones and developmental stages of children, ranging from infants to young adults. For instance, the CDC's Milestone Tracker app aims to give parents a comprehensive overview of children's developmental stages: "*Track your child's milestones from age 2 months to 5 years with CDC's easy-to-use checklists; get tips from CDC to support your child's development; and find out what to do if you ever have concerns about your child's development.*" On the other hand, some apps are more quantitative, such as the Child Growth Tracker app: "*Record multiple children's weight, height, and head circumference measurements and use them to generate growth charts and percentiles from birth to age 20 for some measurements.*"

The five medical services apps allow parents to manage health care services for their children, which can be separated into two subcategories: general medical conditions (four apps) and mental wellbeing (one app). As an example, the app Anytime Pediatrics describes straightforward access to providers: "*Anytime Pediatrics application allows you to see your trusted pediatrician via telemedicine when your child is ill or injured or has a scheduled appointment. Anytime Pediatrics offers you a simple, easy-to-use application and connects you with your pediatric practice.*"

Lastly, there are seven special needs apps that focus on children with learning disabilities, specifically ADHD and Autism. The descriptions clearly state that the apps are developed for individuals diagnosed with learning disorders as shown here: "*AutiSpark is an educational program especially made for children with Autism Spectrum Disorder (ASD).*"

### *Users of the Apps*
To investigate our research question of understanding the user roles, we first analyzed the language used in the app descriptions to determine the expected users of the apps, and afterward, the apps themselves were examined to ascertain the actual users. Our analysis identified two main users involved in the children's health apps: children and parents/caregivers. 30 apps have dual user roles, consisting of both children and parents/caregivers. The rest of the 13 apps have parents/caregivers as the sole user, and we found no apps for children to use exclusively on their own.

*Parents and Children as Dual Users*. After conducting a review of the app features, we discovered that 30 out 43 apps marketed for children have both parents/caregivers and children as dual users. The most common case being the parents have to set up the app and then give the app to children to use. For instance, the hum kids by Colgate app requires the parents/caregivers to create the children's profile and then their own account before giving the app to the children to use. Several apps even directly tell parents when to hand the phone over to their children. In the Habitz: Kids Learn Good Habits app, the parents are prompted to let the children take over after creating an account and setting up the goals: "*Great! Your child is good to go. Let them take it from here.*" The hum kids by Colgate app follows a similar process where they ask the parents/caregivers to create the child's profile and then the child is expected to use the app. The popup speaks to the child: "*Open the app every day to collect your daily rewards.*"

However, it is not clear what is expected of the parents in the app usage. Based on the app descriptions, the children are expected to use the app, but the language used in the app descriptions often addresses the parents/caregivers instead. The messaging is targeted to parents/caregivers, stating the benefits that the children will gain and how they can utilize the app. For example, the Disney Magic Timer by Oral-B tells the parents to "*use this app to seamlessly encourage your kids to brush longer. Longer, happier brushing for your little one is just a download away!*" and the Speech Blubs: Language Therapy app states that it was "*designed to help your child learn new sounds and words, and to practice speaking in a stimulating, educational environment.*" The way the app description is phrased shows how the app developers aim to persuade the parents/caregivers to choose that app for their children: "*KID SAFE & EASY TO USE [...] You can trust your kids are safe with GoNoodle.*" Yet, it is not indicated the extent of the parents/caregivers' roles in the apps other than choosing and setting up the apps.

The misalignment of language, content, and parental controls in the apps also adds to the ambiguity of the expected users. For example, the app, Meditation for Kids, does not require any account creation nor does it have parental controls around settings and purchases. Nonetheless, the name of the app implies that it is for children. It is uncertain if the parents select the videos for the children or if the children are supposed to use the app themselves. The exercise content in the Yoga for Kids & Family Fitness app seems designed for children but it does not offer any parental controls. The language used in the app also addresses the parents/caregivers as shown here: "*Teach your kids with examples - Eating together lets your kids see you eating healthy food and limiting junk food...*" while the content seems to be designed for children. The Language Therapy for Children app further illustrates the prominent role of parents/caregivers: the welcome message is addressed to the parents/caregivers, there is a "Tour for Parents" button that encourages parents/caregivers to get familiar with the app, and a popup before the exercise begins that advises parents to "*please avoid solving puzzles for your child.*" However, it is the children who complete the activities and benefit from the app.

For the apps with parents and children as dual users, we found through app reviews that parents/caregivers tend to express a desire to include children more in the app usage and create a more collaborative use scenario. As children get older, they gain more independence, and parents acknowledge their growth. Parents/caregivers wish to support their children's newfound autonomy in multiple ways, such as giving more device independence, as illustrated in the following quote from a review of a mental health app "*My two older kids each have their own devices and would love to be able to do some of the meditations and stories in their own beds without me having to give up my phone.*" Parents/caregivers also seek to involve more by gradually training them to manage their own health. A reviewer shares her experiences with a period tracker and hopes her daughter will benefit from a menstrual tracking app: "*I have used a period tracker for myself for years and have always found it helpful. I hoped I could find one that would work for my daughter. [...] She doesn't want to use it yet but I know she will.*"

*Parents as Sole Users of Children's Health Apps.* From our analysis, 13 apps did not directly state that the apps were meant to be used by children and instead spoke to the parents/caregivers about how they can use the app to manage their child's health. Though all the apps went through the same screening criteria, 13 apps that resulted are apps that are marketed as children's apps but instead are for parents to manage their kids' health. The language used in the descriptions implies that children do not have any role in the direct usage of the apps. For example, MyCookChildren's mobile app states that the app "*empowers you [parents] to manage all aspects of your child's health while connecting you to the Cook Children's Health Care community.*" One app, Storybook, mentions in the description that "*there are more than 100 stories and massage sessions for children of all ages,*" but after examining the app, the features are designed for the parents/caregivers. The wording in the app speaking to the parents/caregivers indicates that children are not expected to use the app. The parents' role is significant, consisting of account setup and management, tracking their children's data, absorbing parental tips, and following exercises with their children with guidance.

### Goal of Children's Health Apps

Based on the apps' descriptions and features, we found several major goals of the children's health mobile apps: instructional and educational, tracking, introspection, communication, and telehealth services. Many of the apps have multiple goals because they offer a variety of features that help the users achieve those objectives (see Fig. 2.)

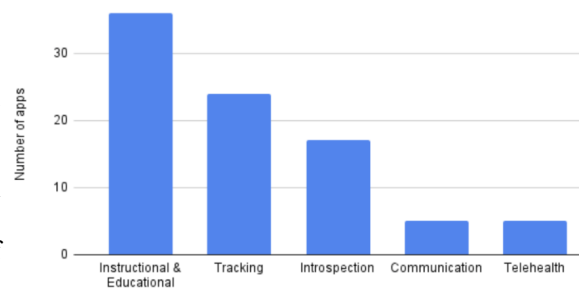

**Fig. 2** Goals of the App

*Instructional and Educational.* Apps that are instructional and educational aim to provide useful information and teach the users health concepts, knowledge, or skills for health management. Educational apps are designed to make the learning process enjoyable and engaging. 36 out of 43 apps aim to be instructional and educational by using methods including storytelling, guidance from instructors, videos, articles, interactive games, and even flashcards. Instructional and educational apps are not limited to one type of user; both children and parents/caregivers can learn from these kinds of mobile apps. However, what children and parents/caregivers learn differs from each other. Parents/caregivers aim to teach children basic health skills such as brushing their teeth or completing chores through the help of the apps. On the other hand, parents/caregivers want to learn about their children's health and look for educational tips on parenting. For example, in the description, the Happy Child-Parenting App "*grants you access to cutting-edge research and tips that will help you raise a happy, well-adjusted child.*"

*Tracking.* Tracking apps allow users to view and manage the overall progression of an event, task, or goal they would like to keep an eye on. Generally, this information is presented through visualization of the data so that it will be easier for people to see progression and patterns. In our study, we found that 24 out of 43 apps utilize progress tracking using features such as checklists and streaks. Other tracking modalities include leaderboards, calendars, report cards, and progress bars. The mobile apps for children offered different tracking functionalities compared to the apps with parents as the sole user. Parents/caregivers generally aim to get a big-picture view of their children's data, and the two major categories are developmental tracking and growth track. The developmental tracking apps mainly use checklists to display progress and the growth tracking apps allow users to input measurements and generate charts that are compared to national percentiles. The majority of the apps offer one or multiple modalities for progress-tracking visualization. To show children their progress, different visualizations are employed: streaks, report cards, and leaderboards.

*Introspection.* Many of the apps analyzed support reflection, and in this study, we use "*introspection*" to label this category of apps. Content such as guided meditations, mindfulness activities, reflection, and breathing exercises all support introspective work. In our analysis, 17 apps were gathered to have introspection as a main function. These apps guide both children and parents, together and separately, through the process of self-reflection. For children, the intent is mainly to be mindful and manage stress. The descriptions state that introspection can be achieved in a variety of ways: "*our mindfulness exercises and stories help children relax, reduce anxiety, and manage negative thoughts*" and "*built on a skills-based approach that helps children in early and middle childhood build a healthier relationship with life, stress, and anxiety…*" For parents/caregivers, the focus is on building the relationship between parent and child. For example, the Storybook app suggests that its app content will help "*sow positive memories and strengthen the relationship between parents and children*" in the description.

*Communication.* We found that six apps focus on supporting communication in managing children's health. Communication-oriented apps offer methods of conversing between users such as between parents and healthcare providers. Five apps have systems where the parents may reach out to providers or nurses via messaging or video. Most of these apps allow bidirectional communication between the parents and providers to discuss the children's health issues; children are not directly involved in the communication. Only one app allows teenagers to directly speak with therapists, and one app enables teenagers to chat with other teenagers in a chatroom open to all users. The majority of these apps provide multiple channels of communication. For instance, the Teen Counseling app supports text messages, video calls, and phone calls to ensure open and accessible contact.

*Telehealth Services.* The last category is telehealth services apps that allow users to access healthcare services and manage overall healthcare records. Five out of 43 apps were identified as telehealth service apps. Common features include scheduling appointments, online consultations for children, ordering prescriptions, accessing immunization records, and even arranging therapy sessions. Telehealth service apps primarily focus on providing access to healthcare services to the parents/caregivers, which is not available in other apps. Only one app allows children to have direct access to a provider.

### Ways to Involve Parents

For the children's mobile health apps that involve both parents/caregivers and children, there is a set of distinct features implemented for parents/caregivers to engage with the app (Fig. 3.) The most prominent features consist of parental controls and account creation. Of the 30 apps designed for dual users, 17 have parental controls, and there are several methods of restriction: using a PIN, entering the birth year, typing in specified numbers, and/or tapping moving objects at the same time. After passing through it, the parents/caregivers have access to profiles, settings, and payment options. Whenever one tries to access the restricted sections, the user would be prompted to "*Grab a parent!*" or "*Verify you are an adult!*" 10 of the 30 required account creation by the parents/caregivers, and only five had the option of creating an account or profile for the children (not necessarily by the children).

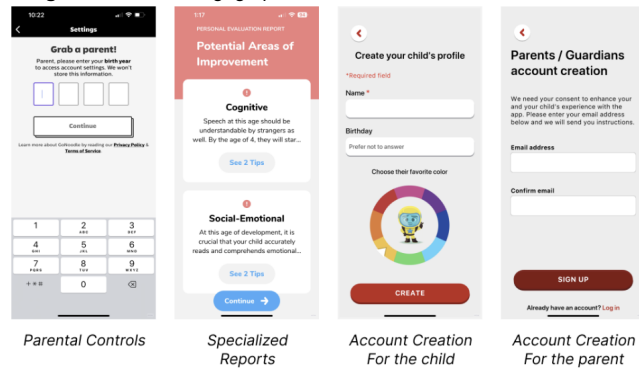

**Fig 3.** Features to engage parents.

The 13 apps specifically designed for parents/caregivers as the sole user focused on a different set of features: account management, telehealth services, educational content, and tracking. All 13 allow parents/caregivers to create profiles for their children and to manage their children's accounts, health, and data through the app. Four apps offer telehealth services including scheduling online appointments with doctors, managing children's health records, and filling prescriptions. These tasks can only be achieved as a parent/caregiver, and parents appreciate the online capabilities: "*With the app everything is so much easier I set up virtual visits. Get the medications I need. Update kids' profiles. Answer questions before the annual doctor visit.*" There are six apps primarily focused on educational content and tracking an aspect of their children's development, which greatly appeal to parents/caregivers. One reviewer value both the education and milestone tracking in one app: "*I love this app been using it since she was about four months old. You input the milestones your child has reached, and it'll give activities to help your child grow…*"

### Ways to Engage Children

Because a larger number of apps consist of dual users, developers employed ways to gain children's attention in the apps. Two key strategies are used to engage children with the apps: interactive media and progress tracking.

All 30 of the apps with dual users offer at least one type of interactive media. Some features include storytelling, following along the instructions from the videos, interacting with cartoon characters, playing games, completing challenges, and solving puzzles. The most notable type of interactive media is videos; 27 apps provide videos for users to follow along. Three apps do not offer videos and instead have storytelling, flashcards for improving verbal communication, and online therapy counseling. These were shown to be effective for some users—one parent praised the app saying that the Card Talk app "*has really opened my son up. […] He is starting to communicate more often &*

*more independently…*" Another review for the Teen Counseling app wrote "*After starting Teen counseling and talking to his counselor, my son started being more open to me about what was going on.*"

Rewards and daily streaks are utilized in the children's health apps to incentivize them by giving tokens to the children in recognition of their work, effort, or achievement. 16 out of 30 apps offer at least one method of rewards or daily streaks, using checklists, streaks, leaderboards, calendars, report cards, and progress bars. The concept of "daily streaks" and "daily rewards" is popular among the children—one parent said in a review that "*my son loves tracking his habits daily and collecting coins*" and another parent mentioned that "*having the multi-day streaks is very motivating.*" The rewards and daily streaks feature motivate the children to use the app for their health, as illustrated in the reviews: "*My kids love it and have been brushing their teeth with no fight now*", "*My kids LOVE this app, ages 4 and 2! It encourages them to brush their teeth and enjoy it too*", and "*My son loves this app! He is very determined by goals and earning things, so this app is perfect for him seeing he's earning coins and working towards a goal…*"

### *Concerns Regarding the App Design*
In this study, we also identified several concerns regarding the design of children's health apps, mainly from the age appropriateness perspective and safety considerations for protecting children's data.

<u>Lack of Specific Age Groups in Children's App.</u> From our evaluation of the app descriptions, we found that most of the apps do not distinguish ages for the users, despite the wide differences among children of different developmental stages. Most apps with children as users use generic language such as "*everything is designed especially for kids*" and "*'Meditation for Kids' is the #1 wellbeing app for kids and is the ultimate happy app for kids.*" Out of the 30 apps, 23 apps are advertised for children of all ages, using general phrases such as "*daily hygiene tracker for kids*" or "*a safe, fun, and entertaining app built just for kids.*" Only 7 out of 30 apps specifically indicated the age groups for the apps, and they varied from 2-5, 3-9, 4-8, 4-10, 7-10, and 3-19+. The app reviews show that children wish for more material based on their age group. One reviewer commented: "*I'm 12 and I have anxiety… the app developer(s) are not really expecting a 10-12-year-old to have trouble sleeping, having anxiety, depression, etc.*" Children's expectations also change over time as they get older, and they realize that the content they used to enjoy when they were younger is not fun anymore: "*that's kinda why I deleted it quite recently… the other reason why is because I'm 10 now! I liked it 3 years ago!*"

<u>Age-inappropriate app content.</u> The app reviews also reveal parents have certain expectations of the apps regarding the content. Because the apps are advertised for children, there is an expectation that the content will be appropriate for children. However, the descriptions may be misleading because many apps have ads, and they may not be child-safe. For example, the Toothy: Teeth Brushing Timer app states that it is a daily hygiene tracker for kids in the description, yet when one review declared that "*the ads that play after the kids' brushing section are NOT for kids*", the developer responded, "*this app does not apply to 'Made for Kids' category and I can't control exactly which advertisements will be shown.*" Another reviewer pointed out the disruptive nature of the ads: "*recently there have been ads popping up right in the middle of these videos. So she's running and jumping and in the middle of swatting and all of the sudden has to abruptly stop, watch the ad, then pick up where she left off. It's super frustrating for her.*" This misdirection can lead to exposure to dangerous content and inappropriate use of the apps.

<u>Protection of Children's Health Data.</u> Another important concern of parents/caregivers is protection, making sure their children and children's data are safe. Since children are especially vulnerable to being taken advantage of, parents/caregivers expect safeguards around their personal information. When apps are not transparent about the reasons for gathering personal data, users become dubious as shown in this review: "*I'm always skeptical what value an account brings to this type of application, especially as I'm very sensitive to my child's privacy. I wish this app had a way to poke around and explore, even if it was just once with no save feature. I am not willing to share any data and chose to delete the app.*" Some features of the app may not be considered safe from the parents/caregivers' perspective: "*Is there a way to make this app safer and make the parental controls work?*"

**Discussion**
In this study, we first discovered that there is a myriad of children's mobile apps that aim to promote health and wellbeing among children. We identified the health focuses and goals of children's mobile health apps, analyzed the primary users and their expectations, and reviewed methods of engagement and involvement implemented. In this section, we discuss the need to explore opportunities to support children with chronic illnesses, allow users to customize health and wellness goals, and design for age appropriateness and digital health safety.

*Designing Mobile Health Apps to Support Chronic Illnesses Management.* Our extensive search on the IOS and Google Play stores revealed that the majority of the apps that fit our criteria for inclusion were wellness apps, such as those for physical activity, sleep, oral hygiene, yoga, and meditation. However, we were unable to include any apps specifically focused on managing chronic illness, despite using targeted keywords (e.g., asthma, cystic fibrosis, diabetes) in our search. This suggests two issues: first, there is a limited selection of commercial apps designed to support children with chronic illness, and second, the apps that do exist for children's chronic health were not widely used. With two in five children aged 6 to 17 in the US living with a chronic health condition[20], it's imperative that more apps are made commercially available to help support those who struggle with these conditions.

While mobile apps have been widely adopted and shown to be effective among adults to manage their health[21–24], our research findings highlight the need for more research and the design of mobile health applications to support children with chronic conditions. Chronic illnesses in children, such as asthma, diabetes, and cystic fibrosis, require continuous management and care. By leveraging the benefits of mobile health apps, such as enhanced medication adherence, health education, self-management, and improved communication with healthcare providers, children with chronic health conditions can be better supported in their journey towards better health and well-being. In addition, mobile apps also offer many fun and engaging features that can motivate and sustain children in their health management. Our results indicate that various wellness apps employ tactics such as gamification, animated videos, and interactive storybooks to boost engagement and education. This implies that future apps designed for children's chronic health management could also utilize these strategies to aid children with chronic health management.

*Supporting Dual Use in Children's Health App.* Our study emphasizes the importance of taking into account both children and caregivers (e.g., parents) when developing a health app for kids. It is crucial to recognize that, even though these mobile apps are intended for children's health, they are never used solely by children. As our findings indicate, most mobile health apps cater to dual use, with parents serving as the primary decision-makers in selecting and configuring the app for their children. App descriptions and promotional materials in app stores predominantly target parents, often overlooking the role and autonomy of children in the app selection. Moreover, we have discovered numerous instances of incongruous app descriptions, which can cause confusion among users and potentially hinder the app's effectiveness. For example, some apps available in app stores are promoted for parents, but they expect children to interact with the app daily. Conversely, certain apps are explicitly marketed for children, yet only parents are anticipated to use them. These observations collectively underscore the necessity to closely examine and support the parent-child relationship and dual-use aspects in order to create a children's app that effectively meets the needs of and empowers both parties.

Future mobile health apps should be developed with dual use in mind, fostering children's involvement in managing their health and wellbeing. Our research shows that parents wish to involve their children more in both app usage and their overall health management. It is crucial to explore ways to enhance children's engagement with health apps. Motivating children to engage with app content and partake in setting their own health goals can create a sense of responsibility and enthusiasm in their app interactions. Furthermore, mobile health apps can be tailored to enable cooperative use between children and caregivers. This may involve reminders or prompts that inspire both the child and caregiver to exercise or assess the child's health progress together. Previous studies have demonstrated that family co-use of health technology can enhance informal health literacy learning and strengthen bonds between children and their family members[25,26]. By fostering collaboration between children and caregivers, mobile health apps can contribute to better health outcomes for children by promoting engagement from both parties.

*Designing for Age-appropriateness and Digital Health Safety.* Through our analysis, we found that most children's health apps do not differentiate children's age groups, despite the significant distinctions between children at various developmental stages[2]. This presents a challenge since a 7-year-old child possesses different cognitive abilities compared to an 11-year-old, let alone a 15-year-old. Children of different ages require different types of content and communication styles. As shown in our results, many users commented that the app provides content that is not appropriate for their age. In addition, some apps are not designed to support older children's need for independence and autonomy. This calls for the need to design mobile apps that provide tailored experiences and content for children of different ages. To ensure comprehensibility, child-friendly language should be used both in the app store and within the app itself, accounting for differences in cognitive and linguistic development across different age groups. In addition, apps need to adopt age-appropriate strategies to engage children. For instance, interactive cartoon videos and storybooks might be more appropriate for younger children than those who are teens. In another example, instructional and health educational materials should be presented in visual, video, or game format for younger children, as many of them have not developed enough linguistic skills to read.

Designing for age appropriateness also calls for special consideration toward children's digital health safety. Our research showed that parents have concerns regarding data protection and the suitability of app content, such as advertisements[27]. To address these worries, three approaches can be taken. Firstly, the policies and community standards for children's data use must be transparent and presented in a comprehensible manner. Apps should clearly state why they need personal (health) information, empowering parents and children to make informed decisions. Secondly, tech companies and policymakers must minimize the negative impact of advertisements in children's digital environment. App stores should enforce strict guidelines for app descriptions and content, preventing developers from falsely promoting their apps as child friendly. Lastly, the safety features in app design must cater to the needs of children of different ages. Providing in-app educational resources is critical to help children and parents develop essential digital and data literacy skills.

**Limitations**
Our study, focused on children's health mobile apps, may not be exhaustive due to the field's vastness and our specific keyword choices. We applied eligibility criteria, included only free apps, and may have overlooked non-English apps or inadvertently included staged reviews. The user reviews we analyzed provide limited, potentially biased insights as they come from users who chose to leave feedback. The full user experience might not be wholly captured through these reviews. Future research could employ methods like interviews or diary studies to explore users' actual experiences with health apps more thoroughly.

**Conclusion**
Our analysis of children's mobile health apps focused on understanding the design and user roles of these technologies. This study identified four prominent health focuses, children and parents as dual users, and methods implemented for app engagement. Our findings indicate that the consumer market lacks children's mobile health apps that support chronic illness management. We also found ambiguity in the expected user roles and the intended age groups. We hope to provide app designers, health researchers, and policymakers with insights and recommendations for promoting children's health and wellbeing through mobile technology.

**Acknowledgements**
This research is funded in part by the UC Irvine Donald Bren School of Information and Computer Sciences Academic Senate Council on Research Computing and Libraries (CORCL) award and the National Science Foundation (NSF) under grant # 2211923.

**References**
1. Mooney CG. Theories of Childhood: An Introduction to Dewey, Montessori, Erikson, Piaget & Vygotsky, Second Edition. Redleaf Press. Redleaf Press; 2013.
2. Piaget J, Inhelder B, Weaver H. The psychology of the child. 1969.
3. Kumar P, Naik SM, Devkar UR, Chetty M, Clegg TL, Vitak J. "No Telling Passcodes Out Because They're Private": Understanding Children's Mental Models of Privacy and Security Online. Proc ACM Hum-Comput Interact. 2017 Dec 6;1(CSCW):64:1-64:21.
4. Santer ND, Manago A, Starks A, Reich SM. Early Adolescents' Perspectives on Digital Privacy. In: Algorithmic Rights and Protections for Children [Internet]. 2021 [cited 2022 Oct 5]. Available from: https://wip.mitpress.mit.edu/pub/early-adolescents-perspectives-on-digital-privacy/release/1
5. Domingues-Montanari S. Clinical and psychological effects of excessive screen time on children. Journal of Paediatrics and Child Health. 2017;53(4):333–8.
6. Saksono H, Castaneda-Sceppa C, Hoffman J, Morris V, Seif El-Nasr M, Parker AG. Storywell: Designing for Family Fitness App Motivation by Using Social Rewards and Reflection. In: Proceedings of the 2020 CHI Conference on Human Factors in Computing Systems [Internet]. New York, NY, USA: Association for Computing Machinery; 2020 [cited 2020 Sep 14]. p. 1–13. (CHI '20). Available from: https://doi.org/10.1145/3313831.3376686
7. Lukoff K, Li T, Zhuang Y, Lim BY. TableChat: Mobile Food Journaling to Facilitate Family Support for Healthy Eating. Proc ACM Hum-Comput Interact. 2018 Nov 1;2(CSCW):114:1-114:28.
8. Pina L, Sien SW, Song C, Ward TM, Fogarty J, Munson SA, et al. DreamCatcher: Exploring How Parents and School-Age Children can Track and Review Sleep Information Together. Proc ACM Hum-Comput Interact. 2020 May 28;4(CSCW1):070:1-070:25.

9. Agapie E, Chang K, Patrachari S, Neary M, Schueller SM. Understanding Mental Health Apps for Youth: Focus Group Study With Latinx Youth. JMIR Form Res. 2022 Oct 18;6(10):e40726.
10. Katule N, Rivett U, Densmore M. A Family Health App: Engaging Children to Manage Wellness of Adults. In: Proceedings of the 7th Annual Symposium on Computing for Development [Internet]. New York, NY, USA: Association for Computing Machinery; 2016 [cited 2021 Nov 27]. p. 1–10. (ACM DEV '16). Available from: https://doi.org/10.1145/3001913.3001920
11. Saksono H, Ranade A, Kamarthi G, Castaneda-Sceppa C, Hoffman JA, Wirth C, et al. Spaceship Launch: Designing a Collaborative Exergame for Families. In: Proceedings of the 18th ACM Conference on Computer Supported Cooperative Work & Social Computing [Internet]. New York, NY, USA: Association for Computing Machinery; 2015 [cited 2020 Sep 14]. p. 1776–87. (CSCW '15). Available from: https://doi.org/10.1145/2675133.2675159
12. Iio M, Miyaji Y, Yamamoto-Hanada K, Narita M, Nagata M, Ohya Y. Beneficial Features of a mHealth Asthma App for Children and Caregivers: Qualitative Study. JMIR mHealth and uHealth. 2020 Aug 24;8(8):e18506.
13. Hong MK, Lakshmi U, Olson TA, Wilcox L. Visual ODLs: Co-Designing Patient-Generated Observations of Daily Living to Support Data-Driven Conversations in Pediatric Care. In: Proceedings of the 2018 CHI Conference on Human Factors in Computing Systems [Internet]. New York, NY, USA: Association for Computing Machinery; 2018 [cited 2021 Nov 27]. p. 1–13. Available from: https://doi.org/10.1145/3173574.3174050
14. Hong MK, Lakshmi U, Do K, Prahalad S, Olson T, Arriaga RI, et al. Using Diaries to Probe the Illness Experiences of Adolescent Patients and Parental Caregivers. In: Proceedings of the 2020 CHI Conference on Human Factors in Computing Systems [Internet]. New York, NY, USA: Association for Computing Machinery; 2020 [cited 2021 Nov 27]. p. 1–16. (CHI '20). Available from: https://doi.org/10.1145/3313831.3376426
15. Potapov K, Marshall P. LifeMosaic: co-design of a personal informatics tool for youth. In: Proceedings of the Interaction Design and Children Conference [Internet]. New York, NY, USA: Association for Computing Machinery; 2020 [cited 2021 Oct 12]. p. 519–31. (IDC '20). Available from: https://doi.org/10.1145/3392063.3394429
16. Hussain MI, Figueiredo MC, Tran BD, Su Z, Molldrem S, Eikey EV, et al. A scoping review of qualitative research in JAMIA: past contributions and opportunities for future work. Journal of the American Medical Informatics Association. 2021 Feb 1;28(2):402–13.
17. Palomba F, Linares-Vásquez M, Bavota G, Oliveto R, Di Penta M, Poshyvanyk D, et al. User reviews matter! Tracking crowdsourced reviews to support evolution of successful apps. In: 2015 IEEE International Conference on Software Maintenance and Evolution (ICSME). 2015. p. 291–300.
18. Torpy JM, Campbell A, Glass RM. Chronic Diseases of Children. JAMA. 2010 Feb 17;303(7):682.
19. Glaser BG. The Constant Comparative Method of Qualitative Analysis. Social Problems. 1965;12(4):436–45.
20. CDC. Healthy Schools [Internet]. Centers for Disease Control and Prevention. 2022 [cited 2023 Mar 21]. Available from: https://www.cdc.gov/chronicdisease/resources/publications/factsheets/healthy-schools.htm
21. Han M, Lee E. Effectiveness of Mobile Health Application Use to Improve Health Behavior Changes: A Systematic Review of Randomized Controlled Trials. Healthc Inform Res. 2018 Jul;24(3):207–26.
22. Mahmood A, Kedia S, Wyant DK, Ahn S, Bhuyan SS. Use of mobile health applications for health-promoting behavior among individuals with chronic medical conditions. DIGITAL HEALTH. 2019 Jan 1;5:2055207619882181.
23. Caldeira C, Chen Y, Chan L, Pham V, Chen Y, Zheng K. Mobile apps for mood tracking: an analysis of features and user reviews. AMIA Annu Symp Proc. 2017;2017:495–504.
24. Su Z, Figueiredo MC, Jo J, Zheng K, Chen Y. Analyzing Description, User Understanding and Expectations of AI in Mobile Health Applications. AMIA Annu Symp Proc. 2021 Jan 25;2020:1170–9.
25. Oygür I, Su Z, Epstein DA, Chen Y. The Lived Experience of Child-Owned Wearables: Comparing Children's and Parents' Perspectives on Activity Tracking. In: In CHI Conference on Human Factors in Computing Systems (CHI '21), May 08–13, 2021, Yokohama, Japan. New York, NY, USA: Association for Computing Machinery; 2021. p. 12.
26. Saksono H, Parker AG. Reflective Informatics Through Family Storytelling: Self-discovering Physical Activity Predictors. In: Proceedings of the 2017 CHI Conference on Human Factors in Computing Systems [Internet]. New York, NY, USA: Association for Computing Machinery; 2017 [cited 2021 Nov 27]. p. 5232–44. (CHI '17). Available from: https://doi.org/10.1145/3025453.3025651
27. Radesky J, Chassiakos YLR, Ameenuddin N, Navsaria D, COUNCIL ON COMMUNICATION AND MEDIA. Digital Advertising to Children. Pediatrics. 2020 Jul;146(1):e20201681.